\documentclass[aip, jap, twocolumn, 10pt, superscriptaddress, showpacs, showkeys, reprint, graphicx]{revtex4-1}
\usepackage{color}
\usepackage{graphicx}
\usepackage{amsfonts}
\usepackage{amsmath}
\usepackage{amssymb}
\usepackage{braket}
\usepackage{mathtools}
\usepackage{siunitx}
\usepackage{tablefootnote}
\usepackage{calc}
\usepackage{verbatim}
\usepackage{enumerate}
\usepackage{textcomp}
\usepackage{scrextend} 
\usepackage[pdftex,
pdfauthor={Matteo Mariantoni}, %
pdftitle={Thin film metrology and microwave loss characterization of indium and aluminum/indium superconducting resonators},
pdfkeywords={Quantum Computing; Superconducting Qubits; Thermocompression Bonding; Indium Thin Films; Sputtering; Thermal Evaporation; Molecular Beam Epitaxy; Secondary-Ion Mass Spectrometry; X-Ray Photoelectron Spectroscopy; Superconducting Planar Resonators; Quality Factor; Loss Tangent; Two-Level States; Interdiffusion; Eutectic}]{hyperref}
\DeclarePairedDelimiterX{\abs}[1]{\lvert}{\rvert}{\ifblank{#1}{{}\cdot{}}{#1}}
\DeclareSIUnit\gauss{G}
\newcommand{\iu}{{i\mkern1mu}}
\begin{document}


\title{\Large{Thin film metrology and microwave loss characterization of indium and aluminum/indium superconducting planar resonators}}

\author{C.R.H.~McRae}
\affiliation{Institute for Quantum Computing, University of Waterloo, 200 University Avenue West, Waterloo, Ontario N2L 3G1, Canada}
\affiliation{Department of Physics and Astronomy, University of Waterloo, 200
University Avenue West, Waterloo, Ontario N2L 3G1, Canada}

\author{J.H.~B\'{e}janin}
\affiliation{Institute for Quantum Computing, University of Waterloo, 200
University Avenue West, Waterloo, Ontario N2L 3G1, Canada}
\affiliation{Department of Physics and Astronomy, University of Waterloo, 200
University Avenue West, Waterloo, Ontario N2L 3G1, Canada}

\author{C.T.~Earnest}
\affiliation{Institute for Quantum Computing, University of Waterloo, 200 University Avenue West, Waterloo, Ontario N2L 3G1, Canada}
\affiliation{Department of Physics and Astronomy, University of Waterloo, 200
University Avenue West, Waterloo, Ontario N2L 3G1, Canada}

\author{T.G.~McConkey}
\affiliation{Institute for Quantum Computing, University of Waterloo, 200
University Avenue West, Waterloo, Ontario N2L 3G1, Canada}
\affiliation{Department of Electrical and Computer Engineering, University of
Waterloo, 200 University Avenue West, Waterloo, Ontario N2L 3G1, Canada}

\author{J.R.~Rinehart}
\affiliation{Institute for Quantum Computing, University of Waterloo, 200
University Avenue West, Waterloo, Ontario N2L 3G1, Canada}
\affiliation{Department of Physics and Astronomy, University of Waterloo, 200
University Avenue West, Waterloo, Ontario N2L 3G1, Canada}

\author{C.~Deimert}
\affiliation{Department of Electrical and Computer Engineering, University of
Waterloo, 200 University Avenue West, Waterloo, Ontario N2L 3G1, Canada}
\affiliation{Waterloo Institute for Nanotechnology, University of Waterloo, 200 University Avenue West, Waterloo, Ontario N2L 3G1, Canada}

\author{J.P.~Thomas}
\affiliation{Department of Electrical and Computer Engineering, University of
Waterloo, 200 University Avenue West, Waterloo, Ontario N2L 3G1, Canada}
\affiliation{Waterloo Institute for Nanotechnology, University of Waterloo, 200 University Avenue West, Waterloo, Ontario N2L 3G1, Canada}

\author{Z.R.~Wasilewski}
\affiliation{Institute for Quantum Computing, University of Waterloo, 200
University Avenue West, Waterloo, Ontario N2L 3G1, Canada}
\affiliation{Department of Physics and Astronomy, University of Waterloo, 200
University Avenue West, Waterloo, Ontario N2L 3G1, Canada}
\affiliation{Department of Electrical and Computer Engineering, University of
Waterloo, 200 University Avenue West, Waterloo, Ontario N2L 3G1, Canada}
\affiliation{Waterloo Institute for Nanotechnology, University of Waterloo, 200 University Avenue West, Waterloo, Ontario N2L 3G1, Canada}

\author{M.~Mariantoni}
\email[Author to whom correspondence should be addressed: ]{\mbox{matteo.mariantoni@uwaterloo.ca}}
\affiliation{Institute for Quantum Computing, University of Waterloo, 200
University Avenue West, Waterloo, Ontario N2L 3G1, Canada}
\affiliation{Department of Physics and Astronomy, University of Waterloo, 200
University Avenue West, Waterloo, Ontario N2L 3G1, Canada}


\date{\today}

\begin{abstract}
Scalable architectures characterized by quantum bits~(qubits) with low error rates are essential to the development of a practical quantum computer. In the superconducting quantum computing implementation, understanding and minimizing materials losses is crucial to the improvement of qubit performance. A new material that has recently received particular attention is indium, a low-temperature superconductor that can be used to bond pairs of chips containing standard aluminum-based qubit circuitry. In this work, we characterize microwave loss in indium and aluminum/indium thin films on silicon substrates by measuring superconducting coplanar waveguide resonators and estimating the main loss parameters at powers down to the sub-photon regime and at temperatures between~$10$ and $\SI{450}{\milli\kelvin}$. We compare films deposited by thermal evaporation, sputtering, and molecular beam epitaxy. We study the effects of heating in vacuum and ambient atmospheric pressure as well as the effects of pre-deposition wafer cleaning using hydrofluoric acid. The microwave measurements are supported by thin film metrology including secondary-ion mass spectrometry. For thermally evaporated and sputtered films, we find that two-level states~(TLSs) are the dominating loss mechanism at low photon number and temperature. Thermally evaporated indium is determined to have a TLS loss tangent due to indium~oxide of~$\sim 5 \times 10^{-5}$. The molecular beam epitaxial films show evidence of formation of a substantial indium-silicon eutectic layer, which leads to a drastic degradation in resonator performance.
\end{abstract}


\maketitle

\section{INTRODUCTION}
    \label{sec:introduction}

The experimental realization of a quantum computer~\cite{Ladd:2010} with hundreds of quantum bits~(qubits), i.e., a medium-scale quantum processor, is on the cusp of becoming a reality. Superconducting quantum computing~\cite{Clarke:2008} has already demonstrated the low error-rate control and measurement of nine qubits~\cite{Kelly:2015} and has all the fundamental attributes required to progress to medium-scale integration in the near future.~\cite{Gambetta:2017:b}

As larger arrays of superconducting qubits become viable, multilayer architectures such as the multilayer microwave integrated quantum circuit~\cite{Brecht:2016} and the three-dimensional integrated quantum processor~\cite{Rosenberg:2017:b} become attractive options for extending current systems. Multilayer architectures are largely composed of two or more on-chip circuits connected by through-silicon vias~\cite{Versluis:2017, Vahidpour:2017} and indium~(In) bump bonds.~\cite{Lewis:2017, Rosenberg:2017:b, Foxen:2017, Versluis:2017, OBrien:2017} Alternatively, pairs of chips can be attached by means of thermocompression bonding of thick film In in ambient atmospheric pressure below the In melting temperature~\cite{Brecht:2017} or thin film In in vacuum above the In melting temperature.~\cite{McRae:2017} Indium is thus becoming an important material to create compact, densely connected, and environment-protected quantum systems. A detailed characterization of loss mechanisms of In thin films is therefore an important step toward a medium-scale quantum processor.

In this article, we study planar superconducting resonators made from In thin films deposited both by thermal evaporation and, separately, grown by molecular beam epitaxy~(MBE), as well as resonators made from sputtered aluminum/indium~(Al/In) thin films. All films are deposited on silicon~(Si) substrates. We find that all devices except for the MBE samples are limited by two-level state~(TLS) loss at the typical excitation power and temperature used in superconducting quantum computing applications. The MBE samples are limited by In-Si eutectic formation and perform significantly worse than all other samples, which, instead, are likely limited by the intrinsic loss due to native In~oxide.

This article is organized as follows. In Sec.~\ref{sec:historical:excursus:and:motivation}, we give a brief historical excursus of the extensive body of work on dissipation in superconducting planar resonators and provide the motivation for this work. In Sec.~\ref{sec:film:deposition:and:fabrication}, we describe the In and Al/In film deposition methods and planar resonator fabrication process. In Sec.~\ref{sec:thin:film:metrology}, we present a detailed characterization of the samples by means of thin film metrology. In Sec.~\ref{sec:resonator:measurements}, we report the quality factor and resonance frequency as a function of both power and temperature for a set of resonators. In Sec.~\ref{sec:results:and:discussion}, we discuss the main results of this work. Finally, in Sec.~\ref{sec:conclusion}, we draw our conclusions and outline possible future work.

\section{HISTORICAL EXCURSUS AND MOTIVATION}
    \label{sec:historical:excursus:and:motivation}

The pursuit of understanding loss mechanisms in thin film technology began in the early stages of superconducting qubit implementations~\cite{Martinis:2005} and has led to major improvements in the quality factor of planar superconducting resonators~\cite{Megrant:2012, Calusine:2017} and coherence time of qubits.~\cite{Barends:2013, Kamal:2016, Gambetta:2017:a} Resonators are particularly amenable to the study of thin film dissipation, which can be quantified from simple transmission-coefficient measurements by estimating the resonator intrinsic (or internal) quality factor~$Q_{\textrm{i}}$,~\cite{Frunzio:2005, Collin:2001} where~$1 / Q_{\textrm{i}} = 1 / Q_{\textrm{c}} + 1 / Q_{\textrm{d}}$. This quantity accounts both for conductor loss~$1 / Q_{\textrm{c}}$ and for dielectric loss~\mbox{$1 / Q_{\textrm{d}} = \tan \delta$}, also known as the~\textit{loss tangent}. Typically, $\tan \delta \simeq \epsilon'' / \epsilon'$, where~$\epsilon'$ and $\epsilon''$ are the real~(lossless) and imaginary~(lossy) part, respectively, of the absolute complex electric permittivity of the dielectric.~\cite{Collin:2001} For superconductors, $1 / Q_{\textrm{c}}$ is determined by effects such as quasiparticles, vortices, metal surface roughness, and radiative losses, while~$\tan \delta$ is determined by effects such as dielectric relaxation and the distribution of TLS defects in the dielectric bulk or surface. The TLSs contribution to~$\tan \delta$ has been identified as one of the dominant extrinsic dissipation channels in superconducting qubits operating at very low temperature ($T \sim \SI{10}{\milli\kelvin}$) and low excitation power (equivalent to a mean photon number~$\langle n_{\textrm{ph}} \rangle \sim 1$).~\cite{Martinis:2005} Under these conditions, the TLSs are unsaturated allowing for the interaction with the qubit states resulting in unwanted dynamics.

The investigation of TLSs in amorphous solids, glasses, and spin glasses at low temperature has occupied a prominent role in condensed matter physics.~\cite{Anderson:1971, Phillips:1981, Phillips:1987} The models developed in those contexts have been adapted to examine loss due to TLSs in on-chip superconducting devices. Coplanar waveguide~(CPW) resonators made from aluminum~(Al) and niobium~(Nb) conductors patterned on various dielectric substrates have been characterized in studies by Gao~\textit{et al.}~\cite{Gao:2007} and Kumar~\textit{et al.},~\cite{Kumar:2008} where TLSs were conjectured to be hosted in either the bulk substrate or native oxide at the substrate-metal~(SM), metal-air~(vacuum)~(MA), and substrate-air~(vacuum)~(SA) interfaces. Experimental evidence for a TLS surface distribution has been later shown in works by Gao~\textit{et al.}~\cite{Gao:2008:a, Gao:2008:b} While a series of studies initially pointed out that TLS loss can be mainly attributed to the MA interface (e.g., native metal oxides),~\cite{Wang:2009, Barends:2010, Sage:2011} it has been later calculated that, for typical conditions, the filling factor~$F$~\cite{Collin:2001} for the SM and SA interfaces can be up to two orders of magnitude larger than that of the MA interface.~\cite{Wenner:2011:b} This suggests that the MA loss~$F_{\textrm{MA}} \tan \delta_{\textrm{MA}}$ dominates the total loss~$F \tan \delta$ only if the intrinsic MA loss~$\tan \delta_{\textrm{MA}}$ is significantly higher than all other intrinsic losses.

The theoretical estimates of Wenner~\textit{et al.}~\cite{Wenner:2011:b} have confirmed some of the findings in a previous work by Wisbey~\textit{et al.},~\cite{Wisbey:2010} where it has been reported that an oxide strip by means of hydrofluoric acid~(HF) of a Si substrate prior to Nb deposition can significantly decrease the total TLS loss~$F \tan \delta_{\textrm{TLS}}$. Further evidence corroborating these results has been presented in the study by Megrant~\textit{et al.},~\cite{Megrant:2012} where the fabrication process of Al CPW resonators was optimized to reduce any SM and SA interface contamination by way of thermal desorption and activated oxygen clean of sapphire substrates in an ultra-high vacuum~(UHV) environment. The same work suggests that the MBE growth of Al on sapphire may also lead to slightly lower loss. Very detailed experiments on loss due to substrate interfaces have been reported both for qubits~\cite{Gambetta:2017:a} and for resonators.~\cite{Calusine:2017} The properties of individual TLSs have been recently studied by Brehm~\textit{et al.}~\cite{Brehm:2017}

The main objective of this work is to characterize the loss mechanisms of In and Al/In CPW resonators on Si substrates. We study resonators with resonance frequency~$f_0 \in ( 4 , 8 ) \, \SI{}{\giga\hertz}$, both at high and low photon number (sub-photon regime) and operated at a temperature~$T \in ( 10 , 450 ) \, \SI{}{\milli\kelvin}$. Assuming
\begin{equation}
\frac{1}{Q_{\textrm{i}} \left( \langle n_{\textrm{ph}} \rangle , T \right) } = F \tan \delta_{\textrm{TLS}} \left( \langle n_{\textrm{ph}} \rangle , T \right) + \frac{1}{Q^{\ast}} \, ,
    \label{eq:1over:Qi:nph:T}
\end{equation}
we estimate~$F \tan \delta_{\textrm{TLS}}$ from the photon number dependence of~$Q_{\textrm{i}}$ at low temperature. Additionally, we estimate the total TLS loss at zero photon number and zero temperature, $F \tan \delta^0_{\textrm{TLS}}$, by fitting the temperature dependence at low photon number of~$1 / Q_{\textrm{i}}$ and $f_0$ to the TLS model of Eqs.~(\ref{eq:F:tan:delta:TLS:T}) and (\ref{eq:Delta:tilde:f:T}), respectively. Finally, we compare the TLS loss to all other losses~$1 / Q^{\ast}$, which we obtain both as a constant offset fitting parameter of the TLS model in Eq.~(\ref{eq:F:tan:delta:TLS:T}) and from~$1 / Q_{\textrm{i}}$ at high photon number.

\section{FILM DEPOSITION AND FABRICATION}
    \label{sec:film:deposition:and:fabrication}

A series of five In and two Al/In films are deposited and patterned for this study. A list of these films and their main features is reported in Table~\ref{table:1}.

\begin{table}[b!]
    \caption{Description of In and Al/In films characterized in this experiment. ``Deposition and metal(s):'' Type of deposition and metal or metals deposited onto sample, in order of deposition. ``Other processing:'' Extra steps performed on sample either before or after deposition. ``Design:'' CPW transmission line design type~$1$ or $2$ (see main text). ``$t_1$; $t_2$:'' Deposited film thickness for Al and In, respectively. ``TE:'' Thermally evaporated film (i.e., unprocessed film). ``Heated:'' Post-patterning heating in ambient atmospheric pressure. ``HF:'' Pre-deposition cleaning with RCA SC-1 and HF dip. ``S:'' Sputtered film (i.e., unprocessed film). ``Heated Vacuum:'' Post-patterning heating in vacuum. ``MBE:'' Molecular beam epitaxial film. ``Annealed:'' Post-deposition annealing in UHV (see main text). Thermally evaporated and sputtered films are deposited on high-resistivity~($> \SI{10}{\kilo\ohm\centi\meter}$) \SI{500}{\micro\meter} thick $3$-in.~float-zone~(FZ) undoped Si~$( 100 )$ wafers; MBE films are deposited on~$3$-in.~Si~$( 001 )$ wafers.}
\begin{center}
    \begin{ruledtabular}
        \begin{tabular}{lccc}
            \raisebox{0mm}[4.5mm][3mm]{\begin{tabular}{@{}c@{}}Deposition \\ and metal(s)\end{tabular}} & \begin{tabular}{@{}c@{}}Other \\ processing\end{tabular} & Design & \begin{tabular}{@{}c@{}}$t_1$; $t_2$ \\ \footnotesize{(\SI{}{\nano\meter})}; \footnotesize{(\SI{}{\nano\meter})}\end{tabular} \\
			\hline
            \raisebox{0mm}[3mm][0mm]{TE In} & \textemdash & $2$ & $0$; $1000$ \\
            \raisebox{0mm}[3mm][0mm]{TE In} & Heated & $2$ & $0$; $1000$ \\
            \raisebox{0mm}[3mm][0mm]{TE In} & HF & $2$ & $0$; $1000$ \\
            \raisebox{0mm}[3mm][0mm]{S Al/In} & \textemdash & $1$ & $150$; $150$ \\
            \raisebox{0mm}[3mm][0mm]{S Al/In} & Heated Vacuum & $1$ & $150$; $150$ \\
            \raisebox{0mm}[3mm][0mm]{MBE In} & \textemdash & $2$ & $0$; $1000$ \\
            \raisebox{0mm}[3mm][0mm]{MBE In} & Annealed & $2$ & $0$; $1000$ \\
            \vspace{-4.5mm}
        \end{tabular}
    \end{ruledtabular}
\end{center}
    \label{table:1}
    \vspace{-3.5mm}
\end{table}

Thermally evaporated In films are deposited in a general-purpose custom thermal evaporator at the Nanotech Nanofabrication Facility of the University of California at Santa Barbara. This evaporator is also used to deposit gold, tin, and other materials with a low melting temperature or high contamination risk. The films are deposited from a~\SI{99.99}{\percent} pure In shot, with a filament voltage between~$20$ and \SI{25}{\volt}, a deposition rate between~$10$ and \SI{15}{\angstrom\per\second}, and a wafer temperature during deposition of less than~$\SI{100}{\degreeCelsius}$. The deposition system is evacuated to~$< \SI{8e-8}{\milli\bar}$ prior to deposition and allowed to cool down after deposition for~$\SI{20}{\minute}$ before venting.

MBE In films are deposited in a system from Veeco Instruments Inc., model GEN10 MBE System at the University of Waterloo. The wafers used for the growth are pre-cleaned by a two-stage outgassing process consisting of a~$\SI{200}{\degreeCelsius}$ anneal in a load-lock followed by a~$\SI{700}{\degreeCelsius}$ anneal in a preparation chamber. An oxide desorption process is conducted by further annealing the wafers in the growth chamber at~$\SI{1040}{\degreeCelsius}$, as measured by a thermocouple.

\begin{figure}[t!]
    \centering
\includegraphics[width=0.99\columnwidth]{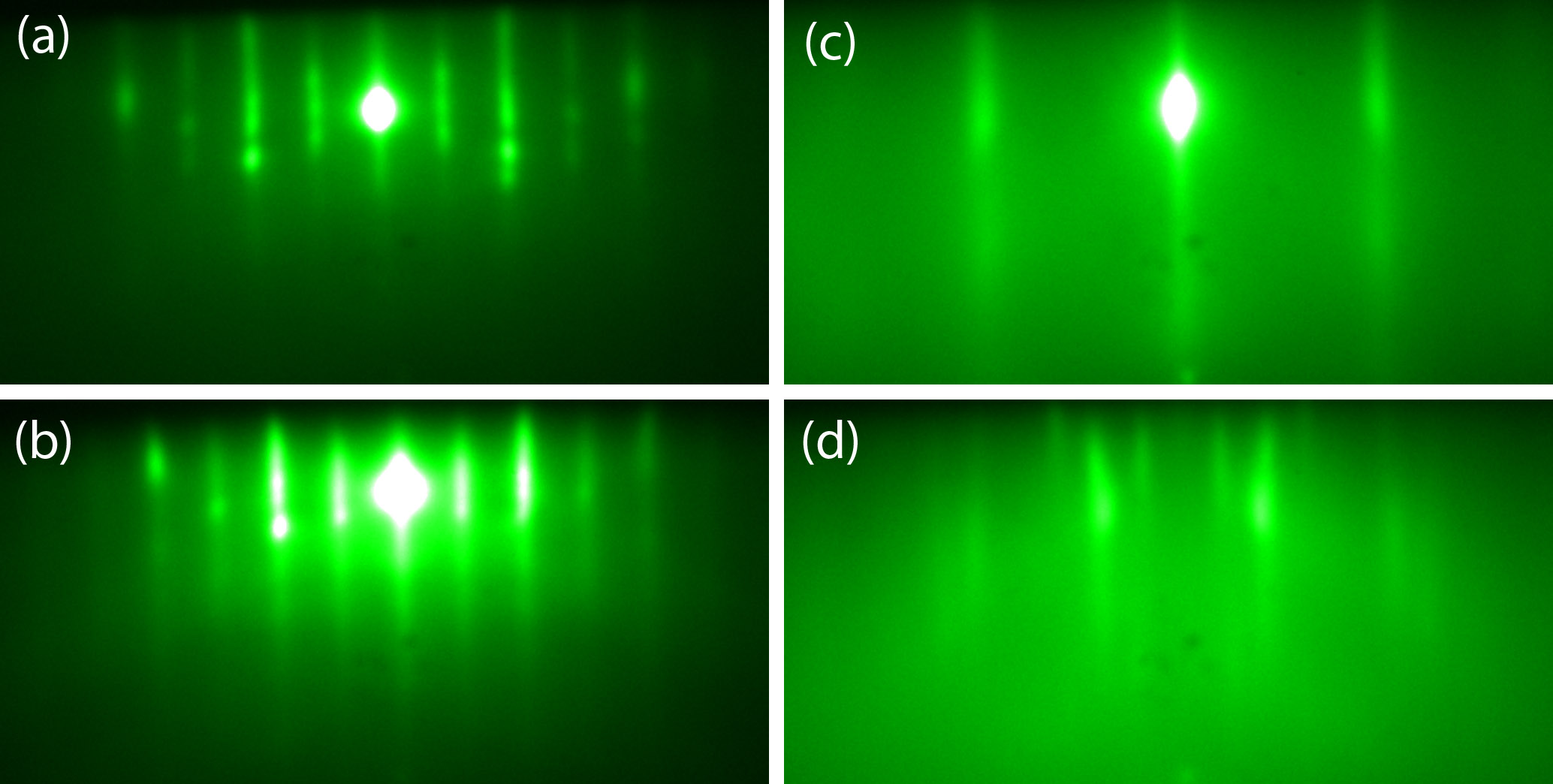}
    \caption{RHEED images. (a) Si wafer after oxide desorption at~\SI{1040}{\degreeCelsius}. (b) Si wafer immediately before starting In deposition at~\SI{0}{\degreeCelsius}. (c) and (d) Immediately post-growth images for the MBE In and MBE In annealed films, respectively. The film in~(d) shows a polycrystalline pattern during growth, with the shown image being captured after additional annealing. The hazy background may indicate the presence of a disordered phase in addition to the single-crystal phase.}
\label{fig:mcrae1ad:main}
\end{figure}

During oxide desorption, the surface reconstruction is monitored by means of an \textit{in situ} reflection high-energy electron diffraction~(RHEED) apparatus comprising a~\SI{12}{\kilo\electronvolt} electron gun from Staib Instruments, Inc., model RHEED-$12$, and a RHEED monitoring system from k-Space Associates, Inc., model kSA~$400$. The latter allows us to capture diffraction images at selected azimuths during wafer rotation. After achieving a sharp and steady $( 2 \times 1 )$ surface reconstruction RHEED pattern [see Fig.~\ref{fig:mcrae1ad:main}~(a)], the wafer is annealed for another~\SI{15}{\minute}, then cooled to~\SI{400}{\degreeCelsius} and kept at this temperature for several hours until the background pressure in the growth chamber drops below~$\sim \SI{2e-10}{\milli\bar}$. The wafer is subsequently ramped to room temperature, at which point the power to the manipulator heater is interrupted and the manipulator is allowed to cool down overnight, resulting in a wafer temperature below~\SI{0}{\degreeCelsius}. This process leads to a mostly atomically clean Si starting surface, as confirmed by the clear Si~$( 001 ) - (2 \times 1 )$ surface reconstruction [see Fig.~\ref{fig:mcrae1ad:main}~(b)].

The In films are deposited at a rate of~$\SI{2}{\angstrom\per\second}$. Film growth is initiated with a wafer temperature below~\SI{0}{\degreeCelsius}; the wafer temperature rises to approximately room temperature during growth as a result of radiative heating from the In effusion cell. Notably, such a deposition temperature is (in Kelvin) more than~\SI{60}{\percent} of the In melting temperature ($\SI{157}{\degreeCelsius}$ at ambient atmospheric pressure). This may lead to significant migration of In on the Si surface not only during growth, but also during storage in UHV. One of the two MBE films is annealed briefly at~\SI{100}{\degreeCelsius} immediately after growth, while still in the MBE chamber (see Table~\ref{table:1}). The samples are kept in UHV overnight before being withdrawn from the MBE system, after which native In~oxide begins to grow on the In film surface, preventing further atom migration and the morphological evolution associated with it. For all samples, no intentional post-growth oxidation is performed.

\begin{figure}[t!]
    \centering
\includegraphics[width=0.99\columnwidth]{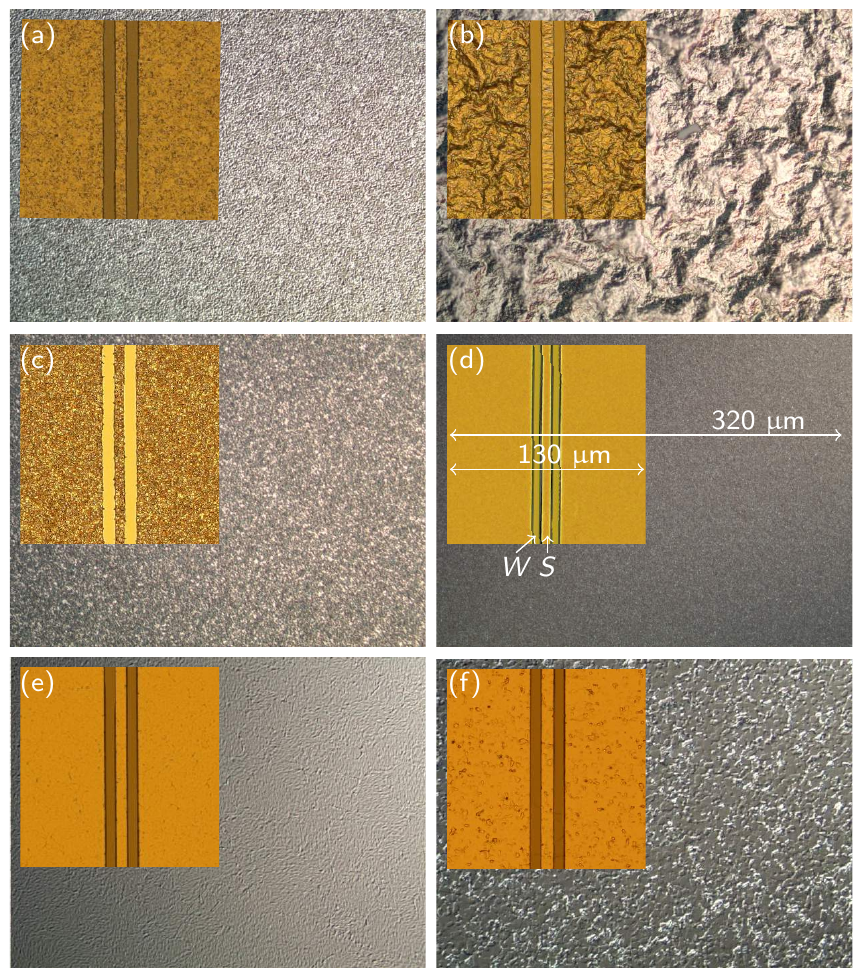}
    \caption{DIC microscopy (background) and standard optical microscopy (insets) of In and Al/In sample surfaces, showing surface and device edge roughness, respectively. Standard microscope images show a CPW transmission line running top to bottom with gaps exposing the Si substrate. Samples shown are thermally evaporated In in~(a), thermally evaporated In heated in~(b), thermally evaporated In HF in~(c), sputtered Al/In in~(d), MBE In in~(e), and MBE In annealed in~(f). Sample details are reported in Table~\ref{table:1}.}
\label{fig:mcrae2af:main}
\end{figure}

CPW transmission line and resonators are defined by optical lithography followed by a wet etch in Transene type~A Al etchant, which successfully etches In as well as Al. Etch times are modified depending on the film type, with thermally evaporated films requiring a wet etch duration of~$\SI{90}{\second}$ and MBE films requiring a shorter etch of~$\SI{60}{\second}$.

After patterning, the thermally evaporated In heated sample is processed by placement on a hot plate at a temperature of~$\SI{190}{\degreeCelsius}$ for~$\SI{5}{\minute}$ in ambient atmospheric pressure.

Prior to film deposition, the thermally evaporated In HF sample is submitted to a cleaning of the Si wafer surface using a~$\SI{15}{\minute}$ ``RCA'' Standard Clean-1~(RCA SC-1) process,~\cite{Kern:1993} immediately followed by removal of the native Si~oxide thin film with a~$\SI{1}{\minute}$ bath in buffered oxide etchant containing~$\SI{1}{\percent}$ HF acid (or \textit{HF dip}). The sample is loaded into vacuum in the thermal evaporator within~\SI{20}{\minute} of the completion of the HF dip.

Sputtered Al/In films are deposited \textit{in situ} in a sputter system from AJA International, Inc., model ATC-Orion 5 at the Toronto Nanofabrication Centre of the University of Toronto (deposition parameters can be found in Ref.~\onlinecite{McRae:2017}).

The sputtered Al/In heated sample is processed in a custom-made vacuum chamber evacuated to~$\SI{1e-2}{\milli\bar}$ that is placed for a time of~$\SI{100}{\minute}$ on a hot plate at~$\SI{190}{\degreeCelsius}$, above the In melting temperature (details on the vacuum chamber in Ref.~\onlinecite{McRae:2017}).

Each film is patterned to form a series of meandered quarter-wave resonators capacitively coupled to a feed CPW transmission line in a multiplexed design (see inset of Fig.~S1 of the supplementary material).~\cite{Bejanin:2016} The resonators feature a center conductor of width~$S$ and gaps of width~$W$, as illustrated in the inset of Fig.~\ref{fig:mcrae2af:main}~(d). For the transmission lines and resonators in design~$1$, $S = \SI{15}{\micro\meter}$ and $W = \SI{9}{\micro\meter}$, and in design~$2$, $S = \SI{12}{\micro\meter}$ and $W = \SI{6}{\micro\meter}$.

Electrical contact to the input and output pads of the feed line occurs through three-dimensional wires.~\cite{Bejanin:2016} Due to the low scratch hardness of In films, we deposit a~$t_2 = \SI{1}{\micro\meter}$ thick In film for the In-only samples to ensure a good electrical connection. In fact, samples featuring a single~$\SI{150}{\nano\meter}$ thick In layer exhibit an exceedingly high contact resistance that makes microwave measurements impossible. The Al/In films, on the other hand, are comprised of a~$t_1 = \SI{150}{\nano\meter}$ thick Al film and a~$t_2 = \SI{150}{\nano\meter}$ thick In film; in this case, the presence of the Al layer guarantees a good electrical connection to the three-dimensional wires.

\section{THIN FILM METROLOGY}
    \label{sec:thin:film:metrology}

\begin{figure*}[t!]
    \centering
\includegraphics[width=0.99\textwidth]{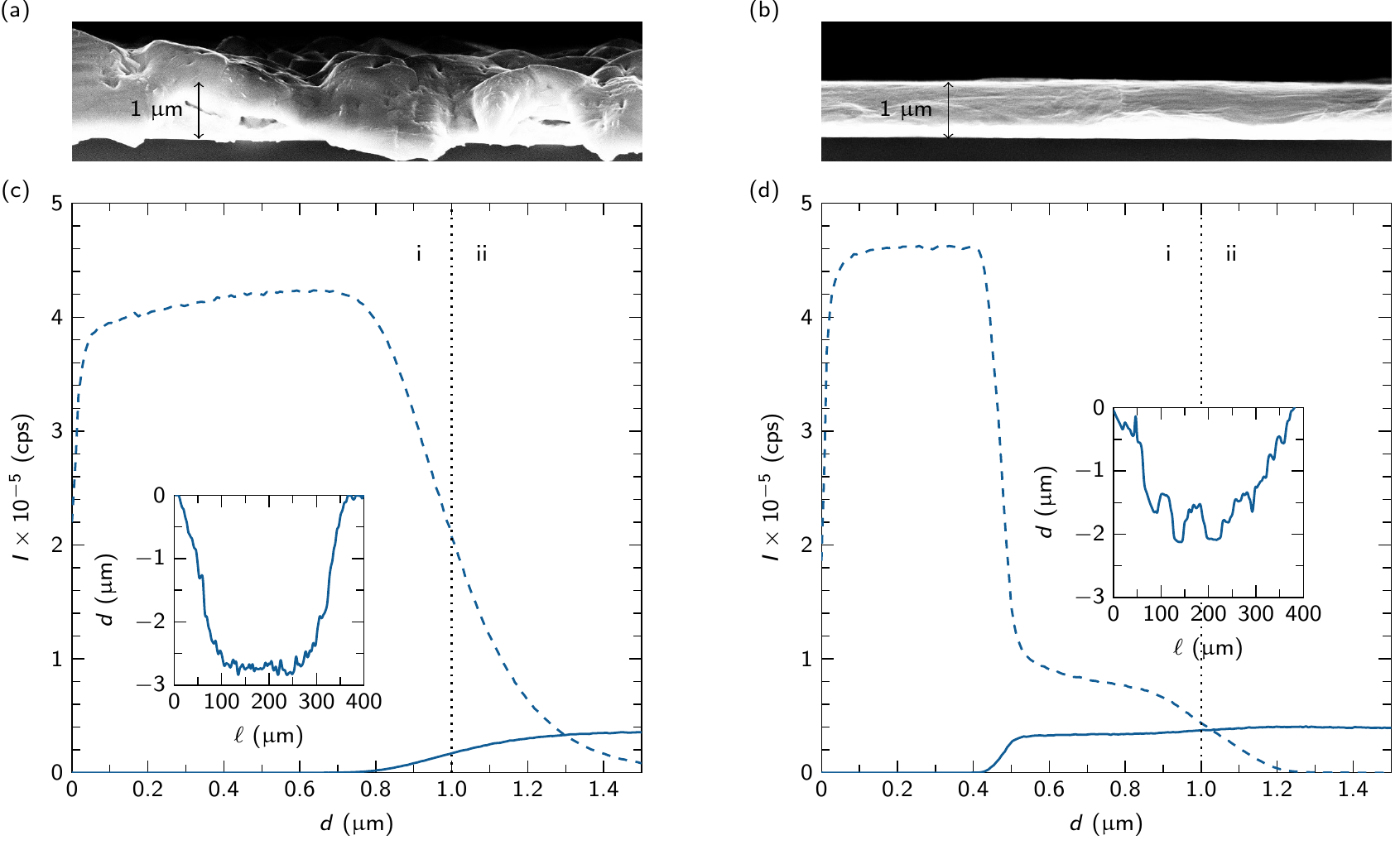}
    \caption{Characterization of In/Si interdiffusion. SEM image of a cleaved thermally evaporated In HF sample~(a) and MBE In sample~(b). D-SIMS depth profiling for the thermally evaporated In HF sample~(c) and MBE In sample~(d) showing measured intensity in counts per second~(cps), $I$, vs.~depth~$d$. Solid lines: Si counts; dashed lines: In counts. Layer~i: $\SI{1}{\micro\meter}$ deep In layer; layer~ii: Top part of Si substrate. Layer separation indicated by vertical dotted black lines. The insets show the D-SIMS crater profile plotted as depth~$d$ vs.~crater diameter~$\ell$.}
\label{fig:mcrae3ad:main}
\end{figure*}

In this section, we study the surface morphology and crystallinity of the samples reported in Table~\ref{table:1} (see Subsec.~\ref{subsec:surface:morphology:and:crystallinity}), Si/In interdiffusion (see Subsec.~\ref{subsec:silicon:indium:interdiffusion}), and surface oxides (see Subsec.~\ref{subsec:surface:oxides}).

\subsection{Surface morphology and crystallinity}
    \label{subsec:surface:morphology:and:crystallinity}

Both differential interference contrast~(DIC) and standard optical microscopy of the surface of the samples in Table~\ref{table:1} (except for the sputtered Al/In heated vacuum sample) are performed, as shown by the images in Fig.~\ref{fig:mcrae2af:main}. DIC microscopy allows the characterization of the surface roughness, whereas standard microscopy is used to verify the smoothness of the main features of CPW lines.

DIC surface microscopy shows extreme roughness on the surface of the thermally evaporated In heated film [see Fig.~\ref{fig:mcrae2af:main}~(b)] and significant roughness on the MBE In film. Roughness on the MBE In film indicates a resemblance to atomically flat insertions blended into a rough, highly textured granular surface. The thermally evaporated In and thermally evaporated In HF films show minor roughness, while the sputtered Al/In and the MBE In annealed films demonstrate very little roughness.

Standard optical microscopy shows signs of roughness on the device edges of the thermally evaporated In HF sample, likely due to the granularity of the film surface itself. Standard microscopy of all other samples shows smooth device edges.

The two MBE In films are extensively characterized throughout the growth by means of \textit{in-situ} RHEED imaging. Pre- and post-growth surface diffraction patterns are displayed in Fig.~\ref{fig:mcrae1ad:main}. The RHEED beam footprint is~$\SI{5}{\milli\meter} \times \SI{0.2}{\milli\meter}$, with an angle of incidence of~$\SI{2}{\degree}$.

RHEED imaging of the MBE In film shows a well-defined~$( 1 \times 1 )$ reconstruction throughout the latter stages of the growth, indicating the presence of a single-crystal phase [see Fig.~\ref{fig:mcrae1ad:main}~(c)]. Measurements of the MBE In annealed film during growth, but before annealing, show a complex RHEED pattern indicative of polycrystalline growth. After annealing in the MBE UHV chamber, RHEED streaks appear, although the background remains hazy and the overall intensity drops [see Fig.~\ref{fig:mcrae1ad:main}~(d)]. This is suggestive of a single-crystal phase coexisting with a disordered phase.

\subsection{Silicon/indium interdiffusion}
    \label{subsec:silicon:indium:interdiffusion}

The interdiffusion of Si and In for the thermally evaporated In HF sample and MBE In sample is characterized by means of scanning electron microscope~(SEM) imaging and dynamic-secondary-ion mass spectrometry~(D-SIMS), as shown in Fig.~\ref{fig:mcrae3ad:main}. SEM allows us to examine a cross section of each sample, while D-SIMS provides information about the layer composition as a function of depth.

SEM images are taken by cleaving a sample and imaging it at a~$\SI{90}{\degree}$ angle, i.e., examining the sample cross section that nominally comprises an In layer above the Si substrate. We use a field-emission~(FE) SEM from Carl Zeiss AG, model LEO FE-SEM~$1530$. All images are taken with a~$\SI{10}{\kilo\volt}$ acceleration voltage. The resulting images are shown in Fig.~\ref{fig:mcrae3ad:main}~(a) and (b). The thermally evaporated In HF film is significantly rougher than the MBE In film, which is extremely smooth. However, the thermally evaporated In HF film is still sufficiently homogeneous to allow for a reliable D-SIMS measurement.

We perform D-SIMS measurements in two different regions of each sample. For all samples, both measurements show similar results. The results for one region of each sample are shown in Fig.~\ref{fig:mcrae3ad:main}~(c) and (d). The samples are analyzed with an ion microprobe from Cameca - AMETEK, Inc., model IMS~$6$f using a positive oxygen beam and monitoring positive secondary ions of interest. The plots show intensity as a function of depth, where the depth scales are obtained by measuring the D-SIMS craters with a surface profilometer from the KLA-Tencor Corporation, model P-$10$ (see insets). There appears to be substantial interdiffusion between In and Si in the thermally evaporated In HF sample.

The profile for the MBE In sample requires a more careful analysis. While there appears to be significant penetration of Si into the In layer, the profile is inconsistent with that of an interdiffusion process. SEM confirms the presence of a~\SI{1}{\micro\meter} In layer on the Si surface, yet D-SIMS shows an abrupt drop in the In count at~$\approx \SI{0.5}{\micro\meter}$ followed by a plateau. This significantly reduced count could be attributed to a change in the SIMS matrix effect, which would in turn indicate an abrupt change in the layer composition and structure. We conjecture that, in fact, an In-Si eutectic of substantial thickness has formed at the interface. The appearance of such a distinct phase would explain the relatively flat SIMS plateaus for both Si and In from~\SI{0.5}{\micro\meter} to \SI{1.0}{\micro\meter}. Furthermore, the abrupt drop in the In count indicates that the crater roughness [see inset of Fig.~\ref{fig:mcrae3ad:main}~(d)] is not present at that point in the sputtering process, but could have developed while sputtering an In-Si eutectic at the interface.

\begin{figure*}[t!]
    \centering
\includegraphics[width=0.99\textwidth]{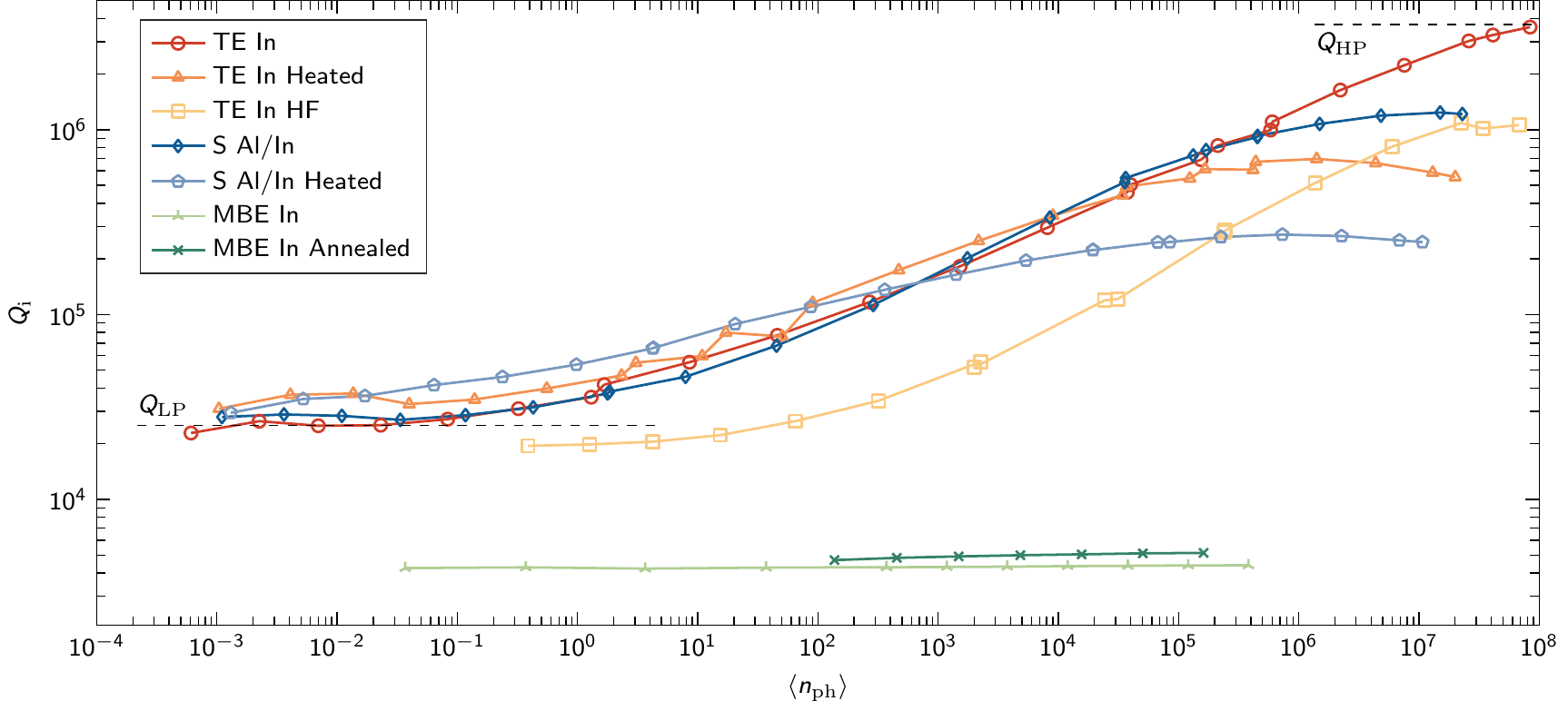}
    \caption{Photon number sweeps (S-curves). Intrinsic quality factor~$Q_{\textrm{i}}$ vs.~mean photon number~$\langle n_{\textrm{ph}} \rangle$ for samples described in Table~\ref{table:1}. The horizontal dashed black lines indicate the estimated values of the high photon number quality factor~$Q_{\textrm{HP}}$ and low photon number quality factor~$Q_{\textrm{LP}}$ for the thermally evaporated In sample. The measurement setup and estimation of~$\langle n_{\textrm{ph}} \rangle$ are the same as in Ref.~\onlinecite{Bejanin:2016}.}
\label{fig:mcrae4:main}
\end{figure*}

\subsection{Surface oxides}
    \label{subsec:surface:oxides}

Surface oxides on all unheated samples and non-annealed MBE sample are measured by means of X-ray photoelectron spectroscopy~(XPS), with measurement results reported in Table~\ref{table:2}. XPS allows for reliable measurements of thin oxide layers in the nanometer range, but not of the thicker oxides expected on heated samples.

\begin{table}[b!]
    \caption{Surface oxide analysis for all unheated samples and non-annealed MBE sample described in Table~\ref{table:1}. ``Sample:'' Sample type (see Table~\ref{table:1}). ``$t_{\textrm{InO}}$:'' Native In~oxide thicknesses; ``$t_{\textrm{SiO}}$:'' Native Si~oxide thicknesses. Two different spots [(A) and (B)] on the Si (gap) surface for the sputtered Al/In sample are measured.}
\begin{center}
    \begin{ruledtabular}
        \begin{tabular}{lcc}
            \raisebox{0mm}[4.5mm][3mm]{Sample} & \begin{tabular}{@{}c@{}}$t_{\textrm{InO}}$ \\ \footnotesize{(\SI{}{\nano\meter})}\end{tabular} & \begin{tabular}{@{}c@{}}$t_{\textrm{SiO}}$ \\ \footnotesize{(\SI{}{\nano\meter})}\end{tabular} \\
            \hline
            \raisebox{0mm}[3mm][0mm]{TE In} & $3.9$ & $0.7$ \\
            \raisebox{0mm}[3mm][0mm]{TE In HF} & $5.1$ & $0.8$ \\
            \raisebox{0mm}[3mm][0mm]{S Al/In} & $4.7$ & $3.4$~(A); $6.5$~(B) \\
            \raisebox{0mm}[3mm][0mm]{MBE In} & $3.1$ & $0.7$ \\
            \vspace{-4.5mm}
        \end{tabular}
    \end{ruledtabular}
\end{center}
    \label{table:2}
    \vspace{-3.5mm}
\end{table}

The samples are analyzed using a spectrometer from Kratos Analytical Ltd, model AXIS Ultra. High-resolution In~$3d$ spectra are obtained from a rectangular spot with dimensions~$\SI{300}{\micro\meter} \times \SI{700}{\micro\meter}$ with a pass energy of~$\SI{10}{\electronvolt}$. High-resolution Si~$2p$ spectra are obtained from a circular spot with diameter~$\SI{110}{\micro\meter}$ with a pass energy of~$\SI{10}{\electronvolt}$; for the Si~$2p$ spectra a region in the CPW gaps is used.

Using a curve of In~oxide thickness as a function of temperature in ambient atmospheric pressure as reported in studies by Kim~\textit{et al.}~\cite{Kim:2008} and Schoeller~\textit{et al.},~\cite{Schoeller:2009} we can estimate the amount of In~oxide on all unheated and MBE samples as~$\approx \SI{5}{\nano\meter}$, confirming the XPS results in Table~\ref{table:2}. The thermally evaporated In heated sample is heated in ambient atmospheric pressure to~$\SI{190}{\degreeCelsius}$, thus growing an estimated~$\SI{20}{\nano\meter}$ layer of In~oxide. The sputtered Al/In heated sample is heated to~$\SI{190}{\degreeCelsius}$, but at a pressure of~$\SI{1e-2}{\milli\bar}$, which likely results in an In~oxide layer thinner than~$\SI{20}{\nano\meter}$ but thicker than~$\SI{5}{\nano\meter}$. Note that the heated samples are particularly hard to measure directly due to the large surface roughness.

\section{RESONATOR MEASUREMENTS}
    \label{sec:resonator:measurements}

In this section, we report~$Q_{\textrm{i}}$ and $f_0$ for comparable resonators on each sample in Table~\ref{table:1}. The circuit layout for sample design~$1$ and $2$ are drawn in the inset of Fig.~S1 of the supplementary material. We present photon number sweeps of~$Q_{\textrm{i}}$ (see Subsec.~\ref{subsec:photon:number:sweeps}) as well as temperature sweeps of~$1 / Q_{\textrm{i}}$ and $f_0$ (see Subsec.~\ref{subsec:temperature:sweeps}), and introduce the TLS theoretical model.

\subsection{Photon number sweeps}
    \label{subsec:photon:number:sweeps}

Transmission-coefficient measurements in the frequency range~$f \in [ 4 , 8 ] \, \SI{}{\giga\hertz}$ at~$T = \SI{10}{\milli\kelvin}$ for each sample in Table~\ref{table:1} are shown in Fig.~S1 of the supplementary material.

Figure~\ref{fig:mcrae4:main} shows~$Q_{\textrm{i}}$ as a function of~$\langle n_{\textrm{ph}} \rangle$ for~$T = \SI{10}{\milli\kelvin}$, where~$Q_{\textrm{i}}$ and $f_0$ are estimated using the fitting procedure explained in Refs.~\onlinecite{Megrant:2012, Bejanin:2016}. The resonators selected for the photon number sweep have resonance frequency at~$T = \SI{10}{\milli\kelvin}$, \mbox{$f_0 \approx \SI{4.387}{\giga\hertz}$} for the thermally evaporated In sample, \mbox{$f_0 \approx \SI{4.377}{\giga\hertz}$} for the thermally evaporated In heated sample, \mbox{$f_0 \approx \SI{4.412}{\giga\hertz}$} for the thermally evaporated In HF sample, \mbox{$f_0 \approx \SI{4.252}{\giga\hertz}$} for the sputtered Al/In sample, \mbox{$f_0 \approx \SI{4.272}{\giga\hertz}$} for the sputtered Al/In heated sample, \mbox{$f_0 \approx \SI{4.800}{\giga\hertz}$} for the MBE In sample, and \mbox{$f_0 \approx \SI{4.790}{\giga\hertz}$} for the MBE In annealed sample. All resonator measurements of thermally evaporated samples correspond to the same designed resonator, as do resonator measurements of sputtered and MBE samples.

At low temperature, where~$k_{\textrm{B}} T \ll h f_0$ ($k_{\textrm{B}}$ and $h$ are the Boltzmann and Planck constant, respectively), the functional dependence of~$1 / Q_{\textrm{i}}$ on~$\langle n_{\textrm{ph}} \rangle$ in the presence of amorphous dielectrics is dictated by TLS saturation above a certain critical mean photon number~$\langle n_{\textrm{ph}} \rangle^{\textrm{c}}$,~\cite{Pappas:2011}
\begin{equation}
F \tan \delta_{\textrm{TLS}} ( \langle n_{\textrm{ph}} \rangle ) \simeq \dfrac{F \tan \delta^0_{\textrm{TLS}}}{\sqrt{1 + \left( \dfrac{\langle n_{\textrm{ph}} \rangle}{\langle n^{}_{\textrm{ph}} \rangle^{\textrm{c}}} \right)^2}} \, .
    \label{eq:F:tan:delta:TLS:nph}
\end{equation}
Thus, we expect to observe a monotonic decrease of~$Q_{\textrm{i}}$ with~$\langle n_{\textrm{ph}} \rangle$, as confirmed by the plots in Fig.~\ref{fig:mcrae4:main}. For high~$\langle n_{\textrm{ph}} \rangle$, $Q_{\textrm{i}}$ reaches a plateau due to the total saturation of the TLSs and where other loss mechanisms dominate, $Q_{\textrm{i}} \simeq Q_{\textrm{HP}}$ (high photon number quality factor). For low~$\langle n_{\textrm{ph}} \rangle$, the curve plateaus at~$Q_{\textrm{i}} \simeq Q_{\textrm{LP}}$ (low photon number quality factor) due to the domination of TLS loss in this region, resulting in an S-shaped curve (or \textit{S-curve}). The term~$1 / Q^{\ast}$ in Eq.~(\ref{eq:1over:Qi:nph:T}) is assumed to be a constant vertical offset of the S-curves.

\subsection{Temperature sweeps}
    \label{subsec:temperature:sweeps}

\begin{figure*}[t!]
    \centering
\includegraphics[width=0.99\textwidth]{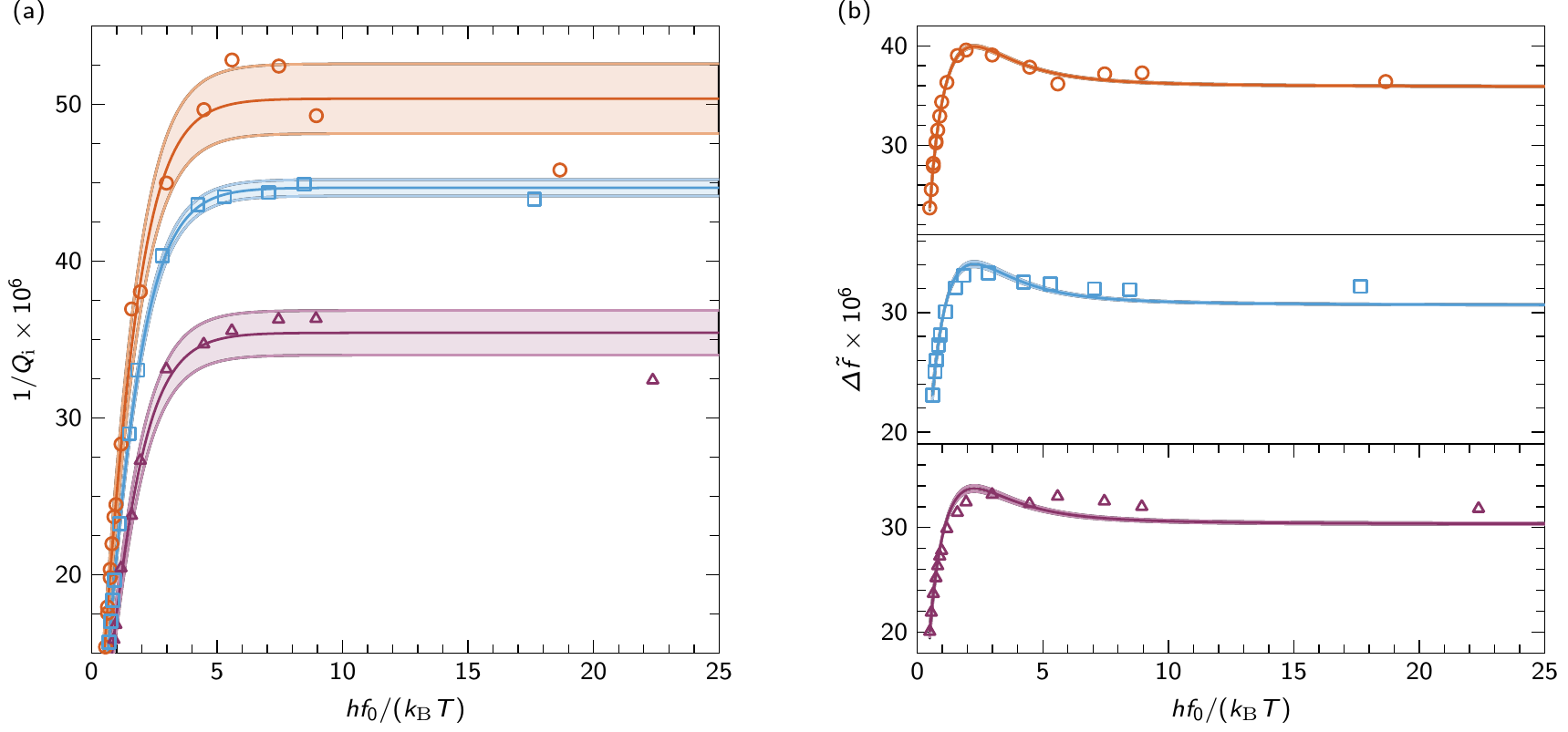}
    \caption{Temperature sweeps. Loss tangent~$1 / Q_{\textrm{i}}$~(a) and normalized frequency shift~$\Delta \tilde{f}$~(b) vs.~$h f_0 / ( k_{\textrm{B}} T )$ for the three thermally evaporated In samples presented in Table~\ref{table:1}. The resonance frequencies at~$T = \SI{10}{\milli\kelvin}$ used in the $x$-axes are reported in Subsec.~\ref{subsec:photon:number:sweeps}. TLS model fitting curves obtained from Eqs.~(\ref{eq:F:tan:delta:TLS:T}) and (\ref{eq:Delta:tilde:f:T}) (solid lines) with standard deviation bands (shaded areas) are overlaid to data points for unprocessed (open circles), heated (open triangles), and HF (open squares) samples.}
\label{fig:mcrae5ab:main}
\end{figure*}

Figure~\ref{fig:mcrae5ab:main} shows the temperature dependence of~$1 / Q_{\textrm{i}}$ and $\Delta \tilde{f}$ at~$\langle n_{\textrm{ph}} \rangle \sim 1$ for all thermally evaporated samples. Similar plots for the sputtered samples is reported in Fig.~S2 of the supplementary material.

At low photon number, $\langle n_{\textrm{ph}} \rangle \sim 1$, the functional dependence of~$1 / Q_{\textrm{i}}$ on~$T$ is due to the interaction between the resonator and TLSs with frequency distribution centered around~$f_0$ (semi-resonant case),~\cite{Pappas:2011}
\begin{equation}
F \tan \delta_{\textrm{TLS}} ( T ) \simeq F \tan \delta^0_{\textrm{TLS}} \, \tanh \left( \dfrac{h f_0}{2 k_{\textrm{B}} T} \right) \, .
    \label{eq:F:tan:delta:TLS:T}
\end{equation}
It can be shown that this relationship is associated with the lossy part of the absolute complex electric permittivity, $\epsilon''$.~\cite{Pappas:2011} We expect to observe a monotonic decrease of~$F \tan \delta_{\textrm{TLS}}$ with~$T$ due to TLS partial saturation activated by thermal photons in the resonator. This behavior is confirmed by the data plotted in Fig.~\ref{fig:mcrae5ab:main}~(a) that was measured up to~$T \approx \SI{450}{mK} \sim T_{\textrm{c}} / 10$, for an In film superconducting transition temperature~$T_{\textrm{c}} = \SI{3.4}{\kelvin}$; under these conditions the quasiparticle contribution to loss is negligible. Also in this case, $1 / Q^{\ast}$ is assumed to be a constant offset. Notably, the data for the sputtered samples reveals quasiparticle loss for~$T \gtrsim \SI{200}{\milli\kelvin}$ due to the lower superconducting transition temperature of the Al film, $T_{\textrm{c}} = \SI{1.2}{\kelvin}$ (see Fig.~S2 of the supplementary material).

Both at low and high photon number, TLSs with frequency distribution largely detuned from~$f_0$ (dispersive case) have almost no contribution to loss. In this case, the TLSs result in a resonator frequency shift given by~\cite{Phillips:1987, Pappas:2011}
\begin{eqnarray}
\lefteqn{ \Delta \tilde{f} ( T ) = \dfrac{f_0 ( T ) - f^0_0}{f^0_0} = } \nonumber \\
& = & \dfrac{F \tan \delta^0_{\textrm{TLS}}}{\pi} \! \left\{ \operatorname{\mathbb{R}e} \left[ \Psi \left( \dfrac{1}{2} + \dfrac{h f_0}{2 \pi \iu k_{\textrm{B}} T} \right) \right] \! - \! \ln \dfrac{h f_0}{k_{\textrm{B}} T} \right\} \!\! ,
    \label{eq:Delta:tilde:f:T}
\end{eqnarray}
where~$f^0_0 = f_0 ( T = 0 )$, $\Psi$ is the complex digamma function, and $\iu^2 = -1$; the values of~$f_0$ used in this equation are reported in Subsec.~\ref{subsec:photon:number:sweeps}. This relationship is associated with the lossless part of the absolute complex electric permittivity, $\epsilon'$.~\cite{Pappas:2011} In this case, we expect a \mbox{non-monotonic} relationship between~$\Delta \tilde{f}$ and~$T$,~\cite{Gao:2008:a, Barends:2010} which is confirmed by the plots in Fig.~\ref{fig:mcrae5ab:main}~(b). To avoid any possible contribution to loss other than TLS loss, these measurements are taken at~$\langle n_{\textrm{ph}} \rangle \sim 1$, although similar results may be obtained at higher photon number.~\cite{Pappas:2011}

\section{RESULTS AND DISCUSSION}
    \label{sec:results:and:discussion}

The resonator measurements shown in Sec.~\ref{sec:resonator:measurements} allow us to estimate:
\begin{enumerate}[I.]
\item $F \tan \delta_{\textrm{TLS}}$ at~$T = \SI{10}{\milli\kelvin}$ as~\cite{Wisbey:2010, Calusine:2017}
\begin{equation}
F \tan \delta_{\textrm{TLS}} \simeq \dfrac{1}{Q_{\textrm{LP}}} - \dfrac{1}{Q_{\textrm{HP}}} \nonumber
    \label{eq:F:tan:delta:TLS}
\end{equation}
and
\begin{equation}
\dfrac{1}{Q^{\ast}} \simeq \dfrac{1}{Q_{\textrm{HP}}} \, ; \nonumber
    \label{eq:1over:Q:ast}
\end{equation}
\item $F \tan \delta^0_{\textrm{TLS}}$ and $1 / Q^{\ast}$ as fitting parameters in Eq.~(\ref{eq:1over:Qi:nph:T}) with~$F \tan \delta_{\textrm{TLS}} ( T )$ given by Eq.~(\ref{eq:F:tan:delta:TLS:T}). This fitting procedure allows us to obtain the fitting curves overlaid to the data in Fig.~\ref{fig:mcrae5ab:main}~(a), which demonstrate a very good agreement with the TLS model;
\item $F \tan \delta^0_{\textrm{TLS}}$ and $f^0_0$ as fitting parameters in Eq.~(\ref{eq:Delta:tilde:f:T}). This fitting procedure allows us to obtain the fitting curves overlaid to the data in Fig.~\ref{fig:mcrae5ab:main}~(b), which also demonstrate a very good agreement with the TLS model.
\end{enumerate}
The estimates for the three thermally evaporated In samples presented in Table~\ref{table:1} are reported in Table~\ref{table:3}. As expected from the design of the samples, the fitted resonance frequencies of the three measured resonators are close to each other, allowing for a fair comparison between different samples. The three estimated values of the TLS loss tangent for each film are in good agreement, demonstrating consistency between different type of measurements and fitting models. A similar argument applies to the two estimated values of other loss mechanisms for each film, where the only significant discrepancy is for the values of the thermally evaporated In film.

At high~$\langle n_{\textrm{ph}} \rangle$, we find that~$1 / Q^{\ast}$ ranges between~\mbox{$\approx 0.3 \times 10^{-5}$} and $\approx 0.3 \times 10^{-6}$ for all devices except for the MBE samples. The MBE resonators are characterized by low performance and display a practically constant~$Q_{\textrm{i}}$ for all values of~$\langle n_{\textrm{ph}} \rangle$. This indicates that the limiting loss mechanism is the presence of an In-Si eutectic phase (see Subsec.~\ref{subsec:silicon:indium:interdiffusion}) rather than TLS loss. The presence of Si in the eutectic possibly results in dielectric relaxation even within the superconducting film.

\begin{table*}[t!]
    \caption{Quantitative analysis of loss mechanisms for the three thermally evaporated In samples. ``$1 / Q_{\textrm{i}} ( \langle n_{\textrm{ph}} \rangle )$:'' S-curve measurements used to estimate~$F \tan \delta_{\textrm{TLS}}$ and $1 / Q_{\textrm{HP}}$. ``$1 / Q_{\textrm{i}} ( T )$ Fit:'' Temperature sweep measurements used to estimate~$F \tan \delta^0_{\textrm{TLS}}$ and $1 / Q^{\ast}$ as fitting parameters. ``$\Delta \tilde{f} ( T )$ Fit:'' Temperature sweep measurements used to estimate~$F \tan \delta^0_{\textrm{TLS}}$ and $f^0_0$ as fitting parameters. See main text for details on fitting models. Each fitting parameter is reported with its standard deviation.}
\begin{center}
    \begin{ruledtabular}
        \begin{tabular}{lcccccc}
            \raisebox{0mm}[4.5mm][3.0mm]{} & \multicolumn{2}{c}{$1 / Q_{\textrm{i}} ( \langle n_{\textrm{ph}} \rangle )$} & \multicolumn{2}{c}{$1 / Q_{\textrm{i}} ( T )$ Fit} & \multicolumn{2}{c}{$\Delta \tilde{f} ( T )$ Fit} \\
            \raisebox{0mm}[3.0mm][0mm]{} & $F \tan \delta_{\textrm{TLS}}$ & $1 / Q_{\textrm{HP}}$ & $F \tan \delta^0_{\textrm{TLS}}$ & $1 / Q^{\ast}$ & $F \tan \delta^0_{\textrm{TLS}}$ & $f^0_0$ \\
            \raisebox{0mm}[3.5mm][0mm]{} & \footnotesize{$\times 10^{-5}$} & \footnotesize{$\times 10^{-5}$} & \footnotesize{$\times 10^{-5}$} & \footnotesize{$\times 10^{-5}$} & \footnotesize{$\times 10^{-5}$} & \footnotesize{(\SI{}{\giga\hertz})} \\
            \hline
            \raisebox{0mm}[4.5mm][0.5mm]{TE In} & $4$ & $0.03$ & $4.70 \mp 0.10$ & $0.33 \mp 0.09$ & $6.1 \mp 0.1$ & $4.665775 \mp 0.000003$ \\
            \raisebox{0mm}[3.5mm][0.5mm]{TE In Heated} & $3$ & $0.20$ & $3.34 \mp 0.08$ & $0.20 \mp 0.06$ & $5.2 \mp 0.3$ & $4.658910 \mp 0.000009$ \\
            \raisebox{0mm}[3.5mm][0.5mm]{TE In HF} & $5$ & $0.09$ & $4.36 \mp 0.03$ & $0.10 \mp 0.02$ & $5.2 \mp 0.3$ & $4.411837 \mp 0.000008$ \\
        \end{tabular}
    \end{ruledtabular}
\end{center}
    \label{table:3}
    \vspace{-3.5mm}
\end{table*}

At low~$\langle n_{\textrm{ph}} \rangle$, we find that all resonators made from thermally evaporated In films perform similarly, following the TLS model with~$F \tan \delta^0_{\textrm{TLS}} \sim 5 \times 10^{-5}$. This behavior persists for the HF dip devices, where the native Si~oxide at the SM interface should be significantly reduced. In all of these devices, the~$F \tan \delta^0_{\textrm{TLS}}$ is approximately five to ten times higher than for the Si/Nb CPW resonators in the study by Wisbey~\textit{et al.},~\cite{Wisbey:2010} suggesting that the intrinsic loss~$\tan \delta_{\textrm{InO}}$ due to native In~oxide at the MA interface is the dominating loss mechanism in all of our In-based resonators. In fact, $\tan \delta_{\textrm{InO}}$ must be large enough to dominate native Si~oxide loss at the SM and SA interfaces, which are characterized by a filling factor significantly larger than the filling factor of the MA interface.~\cite{Wenner:2011:b} It is surprising that devices heated in vacuum and ambient atmospheric pressure, for which the native In~oxide layer at the MA interface is expected to be thicker (see Subsec.~\ref{subsec:surface:oxides}), are also characterized by~$F \tan \delta^0_{\textrm{TLS}} \sim 5 \times 10^{-5}$.

The resonators made from sputtered Al/In films are characterized by a~$Q_{\textrm{i}}$ at low~$\langle n_{\textrm{ph}} \rangle$ on the same order of magnitude as the thermally evaporated resonators. However, the Al/In resonators do not follow the TLS model well, as shown in Fig.~S2 of the supplementary material. This effect may be caused by Si/Al/In interdiffusion, as shown in Fig.~S3 of the supplementary material.

It is worth mentioning that our standard Si/Al resonators are characterized by~$F \tan \delta^0_{\textrm{TLS}} \sim 2 \times 10^{-6}$ at~$\langle n_{\textrm{ph}} \rangle \sim 1$, indicating our setup (with similar features as in Ref.~\onlinecite{Barends:2011}; see also Ref.~\onlinecite{Bejanin:2016}) is adequate to measure ultra-high quality factor resonators.

\section{CONCLUSION}
    \label{sec:conclusion}

In conclusion, we deposit thermally evaporated In, sputtered Al/In, and MBE In films. We characterize the morphology and crystallinity of these films as well their interdiffusion and surface oxides. We fabricate CPW resonators and fit~$Q_{\textrm{i}}$ as a function of~$\langle n_{\textrm{ph}} \rangle$ and $T$ from transmission-coefficient measurements as well as measure~$\Delta \tilde{f}$ as a function of~$T$. We find~$F \tan \delta^0_{\textrm{TLS}} \sim 5 \times 10^{-5}$ and a behavior consistent with TLS dissipation due to the intrinsic loss of native In~oxide for all resonators except for the MBE resonators. The MBE resonators do not follow the TLS model; their substantially reduced resonator performance is consistent with the formation of an In-Si eutectic at the interface. Elucidating this phenomenon will be a subject of a future dedicated study.

Further studies will focus on a more quantitative understanding of the native In~oxide TLS intrinsic loss and on the role of the MA filling factor for In films. In addition, we plan a set of experiments where we vary both the deposition temperature and the deposition rate of thermally evaporated In films after HF dip. In the case of MBE In deposition on Si, the inclusion of an interdiffusion barrier between the Si and In layers (e.g., a very thin titanium nitride layer) may prevent the migration of Si and the formation of an In-Si eutectic, resulting in a substantial increase in the quality factor of the MBE resonators.

In conclusion, our results indicate that In components with exposed or buried In~oxide in a superconducting quantum computer should be limited to the bare minimum and kept far enough from qubits in order to avoid possible qubit degradation.

\section*{SUPPLEMENTARY MATERIAL}

See supplementary material for details on circuit layout and transmission-coefficient measurements, as well as~$1 / Q_{\textrm{i}}$ and $\Delta \tilde{f}$ temperature sweeps and D-SIMS measurements for sputtered Al/In samples.

\begin{acknowledgments}
This research was undertaken thanks in part to funding from the Canada First Research Excellence Fund~(CFREF), the Discovery and Research Tools and Instruments Grant Programs of the Natural Sciences and Engineering Research Council of Canada~(NSERC), the Ministry of Research and Innovation~(MRI) of Ontario, and the Alfred P.~Sloan Foundation. We would like to acknowledge the Canadian Microelectronics Corporation~(CMC) Microsystems for the provision of products and services that facilitated this research, including CAD software and HFSS, as well as the Quantum NanoFab Facility at the University of Waterloo. We thank Tom Reynolds and Biljana Stameni\'{c} at the Nanotech Nanofabrication Facility of the University of California at Santa Barbara for depositing the thermally evaporated In films, Harlan Kuntz and Edward Huaping Xu at the Toronto Nanofabrication Centre of the University of Toronto for depositing the sputtered Al/In films, as well as Gary Good and Mark C.~Biesinger at the Surface Science Western laboratory of the University of Western Ontario for their help with SIMS and XPS measurements. We thank Adel O.~Abdallah for assisting during a sample heating, and acknowledge our fruitful discussions with Robert F.~McDermott and Guo-Xing Miao.
\end{acknowledgments}

\clearpage

\pagebreak

\newcommand{\beginsupplement}{%
    \setcounter{section}{0}
    \renewcommand{\thesection}{S\Roman{section}}%
    \setcounter{subsection}{0}
    \renewcommand{\thesubsection}{\Alph{subsection}}%
    \setcounter{subsubsection}{0}
    \renewcommand{\thesubsubsection}{S\Alph{subsubsection}}%
    \titleformat{\subsubsection}[block]{\bfseries\centering}{\thesubsubsection.}{1em}{}
    \setcounter{table}{0}
    \renewcommand{\thetable}{S\Roman{table}}%
    \setcounter{figure}{0}
    \renewcommand{\thefigure}{S\arabic{figure}}%
    \setcounter{equation}{0}
    \renewcommand{\theequation}{S\arabic{equation}}%
    }

\setcounter{page}{1}
\makeatletter

\renewcommand{\bibnumfmt}[1]{[S#1]}
\renewcommand{\citenumfont}[1]{S#1}

\section*{Supplementary material for ``Thin film metrology and microwave loss characterization of indium and aluminum/indium superconducting planar resonators''}

This supplementary material is organized as follows. In Sec.~\ref{sec:circuit:layout:and:transmission:coefficient:measurements}, we provide further details on circuit layout and transmission-coefficient measurements. In Sec.~\ref{sec:temperature:sweeps:of:aluminum:indium:resonators}, we show~$1 / Q_{\textrm{i}}$ and $\Delta \tilde{f}$ temperature sweeps for sputtered Al/In samples. Finally, in Sec.~\ref{sec:silicon:aluminum:indium:interdiffusion} we present dynamic-secondary-ion mass spectrometry~(D-SIMS) measurements for sputtered Al/In samples, showing Si/Al/In interdiffusion.

\section{CIRCUIT LAYOUT AND TRANSMISSION-COEFFICIENT MEASUREMENTS}
    \label{sec:circuit:layout:and:transmission:coefficient:measurements}

\begin{figure*}
    \centering
\includegraphics[width=0.99\textwidth]{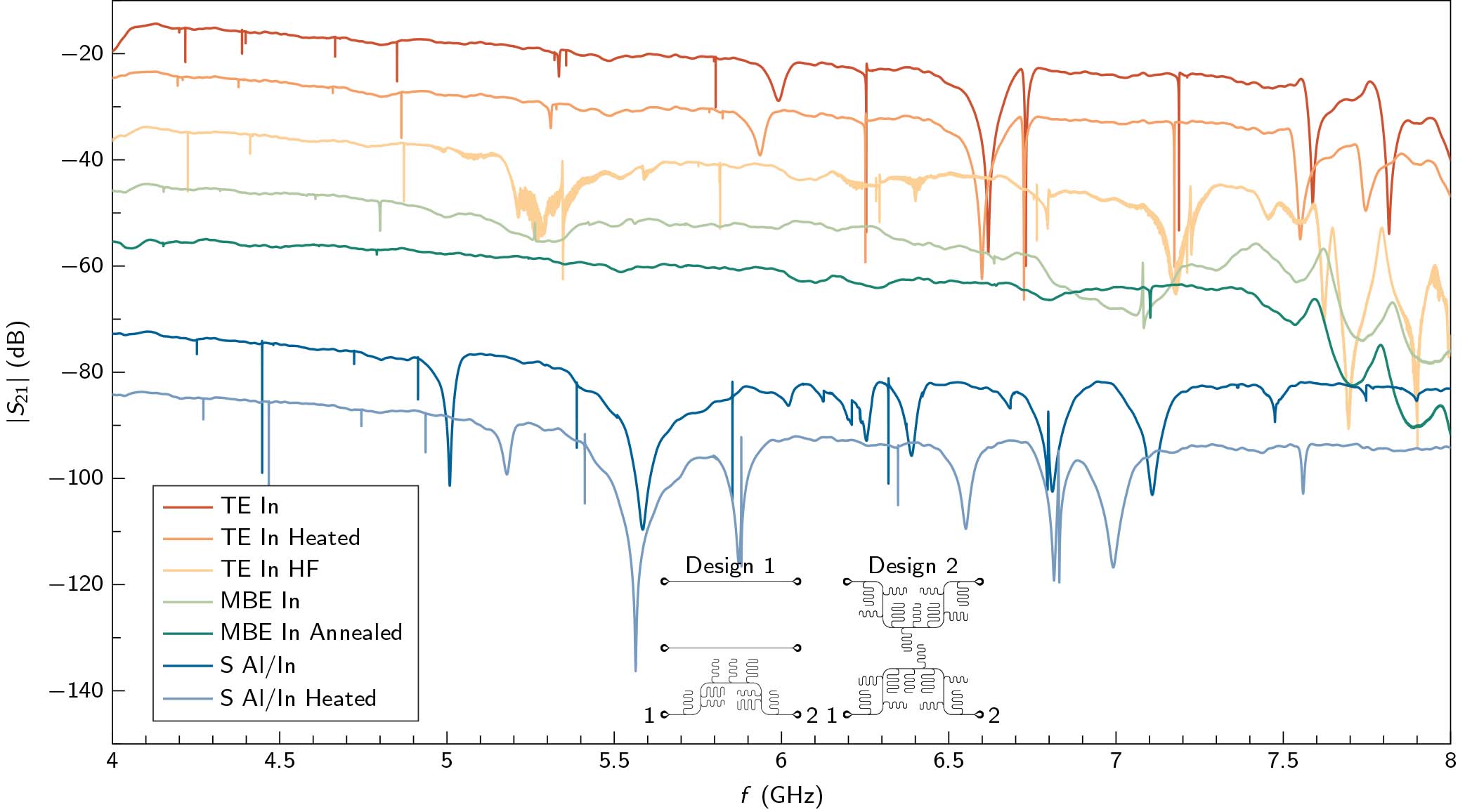}
    \caption{Transmission-coefficient measurements. Magnitude of the measured
    transmission coefficient~$| S_{21}|$ vs. frequency~$f$. Inset: Circuit layout of design~$1$ (left) and design~$2$ (right). In both layouts, the measured CPW transmission lines are those between ports~$1$ and $2$.}
\label{fig:mcrae1:suppl}
\end{figure*}

For the samples with design~$1$ (see Sec.~III of the main text for details), the circuit layout comprises a set of nine quarter-wave resonators capacitively coupled to the coplanar waveguide~(CPW) transmission line between ports~$1$ and $2$ in a multiplexed design, as shown in the inset of Fig.~\ref{fig:mcrae1:suppl}. The other two CPW transmission lines in this layout are not used in this work.

For the samples with design~$2$, the circuit layout comprises two sets of ten quarter-wave resonators, with one set of resonators capacitively coupled to the CPW transmission line between ports~$1$ and $2$, also in a multiplexed design, as shown in the inset of Fig.~\ref{fig:mcrae1:suppl}. The second CPW transmission line in this layout is not used in this work.

Transmission-coefficient measurements in the frequency range~$f \in [ 4 , 8 ] \, \SI{}{\giga\hertz}$ at~$T = \SI{10}{\milli\kelvin}$ for each sample in Table~I of the main text are shown in Fig.~\ref{fig:mcrae1:suppl}.

The measurements in Fig.~\ref{fig:mcrae1:suppl} demonstrate a stark difference in performance between resonators on thermally evaporated In samples and the two MBE In samples. Out of the ten designed resonators, seven are successfully detected and fitted for thermally evaporated In samples, while only one and two are found for the MBE In and MBE In annealed samples, respectively.

We note that some of the~$| S_{21} |$ traces in Fig.~\ref{fig:mcrae1:suppl} show the presence of unwanted modes, particularly for the sputtered samples. These modes are probably slotline modes due to broken ground planes.~\cite{Wenner:2011:a:s}

\section{TEMPERATURE SWEEPS OF ALUMINUM/INDIUM RESONATORS}
    \label{sec:temperature:sweeps:of:aluminum:indium:resonators}

\begin{figure*}
    \centering
\includegraphics[width=0.99\textwidth]{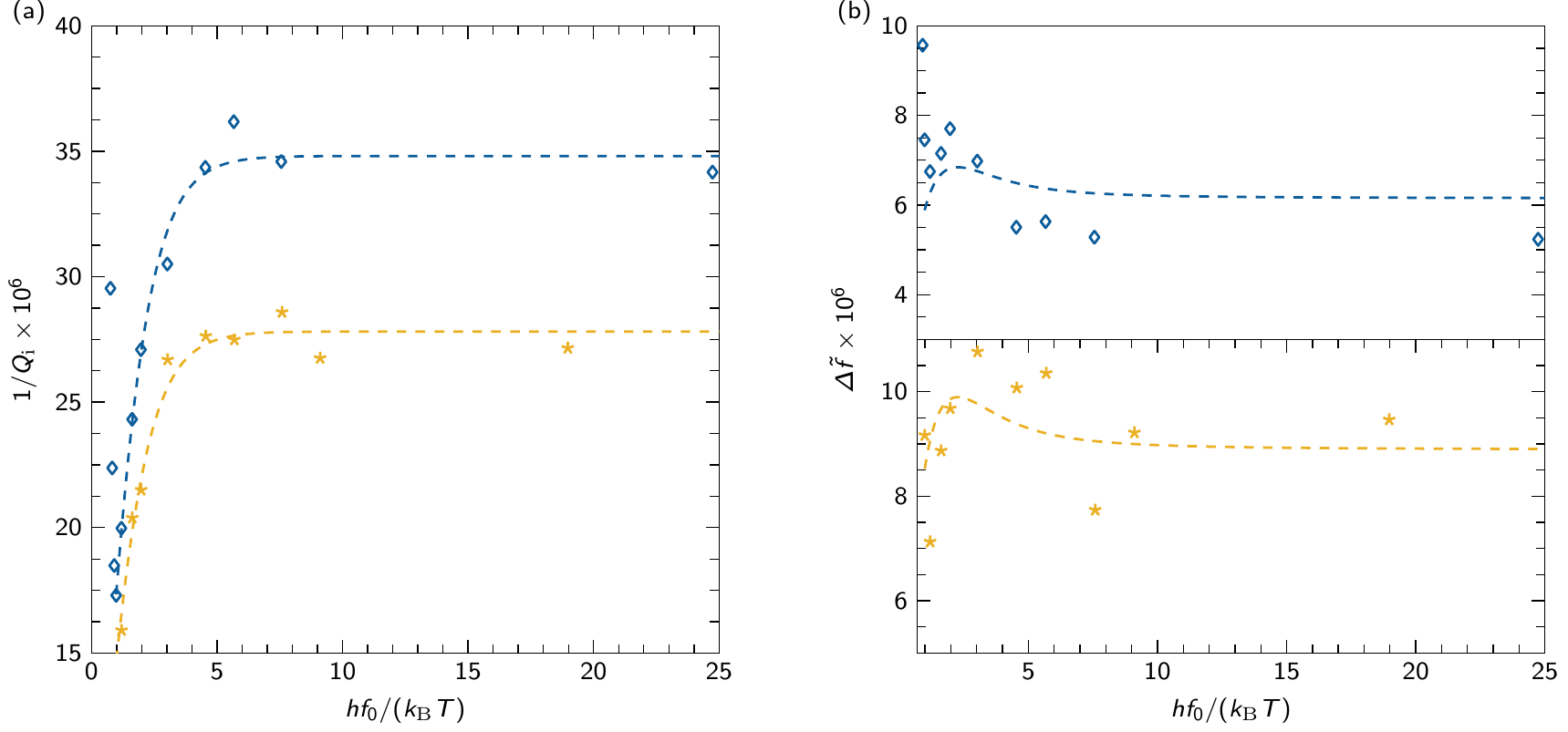}
    \caption{Temperature sweeps. Loss tangent~$1 / Q_{\textrm{i}}$~(a) and normalized frequency shift~$\Delta \tilde{f}$~(b) vs.~$h f_0 / ( k_{\textrm{B}} T )$ for the two sputtered Al/In samples presented in Table~I of the main text. The resonance frequencies at~$T = \SI{10}{\milli\kelvin}$ used in the $x$-axes are reported in Subsec.~VA of the main text. TLS model fitting curves (dashed lines) are overlaid to data points for unprocessed (open diamonds) and heated (stars) samples.}
\label{fig:mcrae2ab:suppl}
\end{figure*}

Figure~\ref{fig:mcrae2ab:suppl} shows the temperature dependence of~$1 / Q_{\textrm{i}}$ and $\Delta \tilde{f}$ at~$\langle n_{\textrm{ph}} \rangle \sim 1$ for the two sputtered Al/In samples. The data is overlaid with fitting curves obtained using the TLS theoretical model of Eqs.~(1), (3), and (4) of the main text.

The sputtered Al/In samples do not fit well to the TLS model, possibly due to loss caused by significant interdiffusion of the In and Al layers and Si substrate (see Sec.~\ref{sec:silicon:aluminum:indium:interdiffusion}).

\section{SILICON/ALUMINUM/INDIUM INTERDIFFUSION}
    \label{sec:silicon:aluminum:indium:interdiffusion}

\begin{figure*}
    \centering
\includegraphics[width=0.99\textwidth]{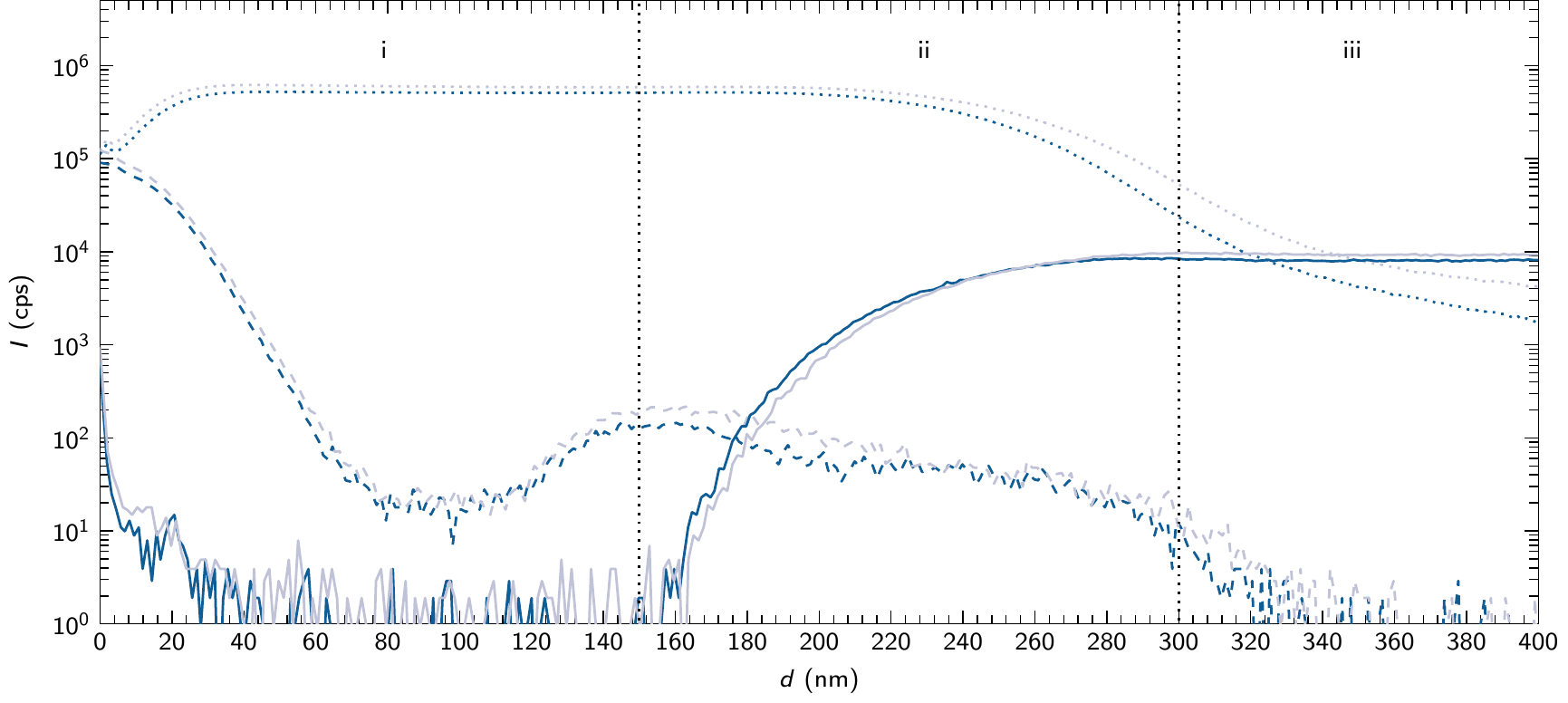}
    \caption{Characterization of Si/Al/In interdiffusion. D-SIMS depth profiling for the sputtered Al/In sample~[dark blue (dark gray)] and sputtered Al/In heated sample~[light blue (light gray)] showing measured intensity in counts per second~(cps), $I$, vs.~depth~$d$. Dashed lines: In counts; dotted lines: Al counts; solid lines: Si counts. Layer~i: $\SI{150}{\nano\meter}$ deep In layer; layer~ii: $\SI{150}{\nano\meter}$ deep Al layer; layer~iii: Top part of Si substrate. Layer separation indicated by vertical dotted black lines.}
\label{fig:mcrae3:suppl}
\end{figure*}

Figure~\ref{fig:mcrae3:suppl} shows D-SIMS measurements of the two sputtered Al/In samples in Table~I of the main text. The instrument details can also be found in the main text. The results indicate significant diffusion of Si into the Al layer, possibly leading to dielectric relaxation within the Al layer. In addition, we notice extreme interdiffusion of Al into In up to the surface of the In layer. The sputtered Al/In heated sample shows a slightly higher level of Al diffusion into the Si surface, but otherwise the samples have similar D-SIMS profiles.

\newpage

\end{document}